\documentclass[a4paper]{article}

\usepackage{fullpage} 
\usepackage{pstricks}
\usepackage{multicol}
\usepackage{graphicx}
\usepackage{movie15}
\usepackage{hyperref}
\usepackage[version=3]{mhchem}
\usepackage{amsmath} 
\usepackage{amsfonts}
\usepackage{color}
\usepackage{soul}
\usepackage[margin=1.7cm,top=3cm,bottom=2.8cm,centering]{geometry}

\definecolor{rosepale}{rgb}{1.0, 0.7, 1.0}

\newcommand{\be}{\begin{equation}}
\newcommand{\ee}{\end{equation}}
\newcommand{\bea}{\begin{eqnarray}}
\newcommand{\eea}{\end{eqnarray}}

\title{The accuracy of biochemical interactions is ensured by endothermic stepwise kinetics\footnote{Reference: Michel D, Boutin B, Ruelle P. 2016. The accuracy of biochemical interactions is ensured by endothermic stepwise kinetics. Prog. Biophys. Mol. Biol. 121, 35-44.}}

\author{Denis Michel, Benjamin Boutin $ \dagger $ and Philippe Ruelle $ \ddagger $ \\
\\
      \begin{small} Universite de Rennes1-IRSET. Campus Sant\'e de Villejean. 35000 Rennes France. denis.michel@live.fr \end{small} \\ \begin{small} $ \dagger $ Universite de Rennes1-Institut de Recherche Math\'ematiques de Rennes (IRMAR)\end{small} \\ \begin{small} Campus de Beaulieu Bat. 22/23.  35042 Rennes France  \end{small} \\ \begin{small} $ \ddagger $ Universit\'e catholique de Louvain - UCL. Institut de Recherche en Math\'ematique et Physique \end{small} \\ \begin{small} IRMP. Chemin du Cyclotron, 2. B-1348 Louvain-la-Neuve, Belgium.\end{small}
\\}

\date{} 

\begin{document}
\maketitle

\begin{multicols}{2}

\section{Abstract}
The discerning behavior of living systems relies on accurate interactions selected from the lot of molecular collisions occurring in the cell. To ensure the reliability of interactions, binding partners are classically envisioned as finely preadapted molecules, evolutionarily selected on the basis of their affinity in one-step associations. But the counterselection of inappropriate interactions can in fact be much more efficiently obtained through difficult multi-step adjustment, whose final high energy state is locked by a fluctuation ratchet. The progressive addition of molecular bonds during stereo-adjustment can be modeled as a predominantly backward random walk whose first arrival is frozen by a micro-irreversible transition. A new criterion of ligand specificity is presented, that is based on the ratio rejection/incorporation. In addition to its role in the selectivity of interactions, this generic recipe can underlie other important biological phenomena such as the regular synthesis at low level of supramolecular complexes, monostable kinetic bimodality, substrate concentration thresholds or the preparation of rapidly depolymerizable structureswith stored energy, like microtubules. \\
\newline
\textit{Keywords}: Fluctuation ratchet; self-assembly; substrate selection; binding specificity; induced fit.

\section{Introduction}

Macromolecular crowding is the rule in most cellular compartments, but only certain interactions are appropriate, which imposes stringent partner selection. The preference for appropriate over inappropriate interactions, is classically assumed to rely on optimal conformational preadjustment between co-evolved complementary macromolecules. This type of binding is exothermic, that is to say thermodynamically driven by stabilization, which can be monitored in microcalorimetry by a dissipation of heat. But beside this standard mode of binding, authors understood that other mechanisms should exist to discriminate closely related, wrong and correct substrates. This discernment is necessary for example in the case of polymerases which should accommodate different substrate molecules at each polymerization step \cite{Hopfield,Ninio1975}. To ensure the counterselection of undesired substrates, the activity of theses polymerases should be low enough and the probability of substrate dissociation relatively high. By this way, slight differences of dissociation rates are amplified and more opportunities are given to inappropriate substrates to leave the enzyme before incorporation \cite{Ninio1975}. Authors then showed that increasing the number of proofreading steps with irreversible ligand exit, can strikingly decrease the error rate \cite{Savageau,Blomberg}. In these studies, the successive rounds of substrate checking are fundamentally driven and energy-consuming, in line with the expected thermodynamic cost of accuracy. This property is however not necessary if spontanous thermal fluctuations can be exploited. The classical one-step lock-and-key binding can be stabilized by induced fit, but the importance of conformational adjustment is variable. Contrary to initial binding that is a single step, conformational adjustment is multistep and can lead to highly selective interactions from moderately pre-adjusted macromolecules. Moreover, inefficient adjustment is shown here capable of amplifying slight advantages of desired interactions over nonrelevant ones, as quantified by the ratio rejection/incorporation. This mechanism is modeled as a chain of reversible events locked in its final state by a micro-irreversible transition, similar to a random walk with a final absorbing state. Different applications of this general principle in biochemical systems are presented.

\section{Multistep interactions}

Protein binding is a multifaceted mechanism including a diffusion step, initial binding and conformational adaptation \cite{Zhou}. The two latter processes are generally compressed into a single one, to give

\begin{subequations} \label{E:gp}  
\begin{equation} k_{+} = \dfrac{k_{\textup{on}}k_{\textup{i}}}{k_{\textup{off}}+k_{\textup{i}}} \end{equation} \label{E:gp1}
\noindent
and
\begin{equation} k_{-} = \dfrac{k_{\textup{off}}k_{-\textup{i}}}{k_{\textup{on}}+k_{-\textup{i}}} \end{equation} \label{E:gp2}
\end{subequations} 
\noindent
where $ k_{\textup{on}} $ is a pseudo-first order rate including the component concentration and $ k_{i} $ is the rate of conformational adjustment. It is shown below that very interesting behaviors emerge if this conformational adaptation is itself not a single step but a series of micro-reactions. In the induced-fit mode of interaction, the molecular partners first interact incompletely and then  progressively adapt to one another \cite{Koshland}. This conformational moulding can be driven by an increase of overall stability through the formation of additional bonds, but less intuitively, it can also be counter-thermodynamic and obtained by chance if the addition of intermolecular bonds is opposed to a conformational resistance. This latter possibility can become significant if the most adjusted complex is ratified by a final irreversible step, whose nature depends on the system considered. This general mechanism is illustrated below in the context of the self-assembly of supramolecular complexes and of enzyme-substrate interactions.  

\subsection{Hierarchical addition of components in molecular complexes}

Essential cellular functions are ensured by multimolecular complexes made of many individual components, essentially proteins, but also sometimes structural RNAs, whose assembly can be assisted by helper proteins (chaperones) and is generally ordered, at least by parts \cite{Chari} (Fig.1A). Microreversible adjustement chains can have different thermodynamic properties, schematized in Fig.2. They are traditionally considered as thermodynamically favored and associated with a decrease of energy (Fig.2A), but in fact nothing forbids counter-thermodynamic binding to occur and to be stabilized (Fig.2B).

\begin{center}
\includegraphics[width=7cm]{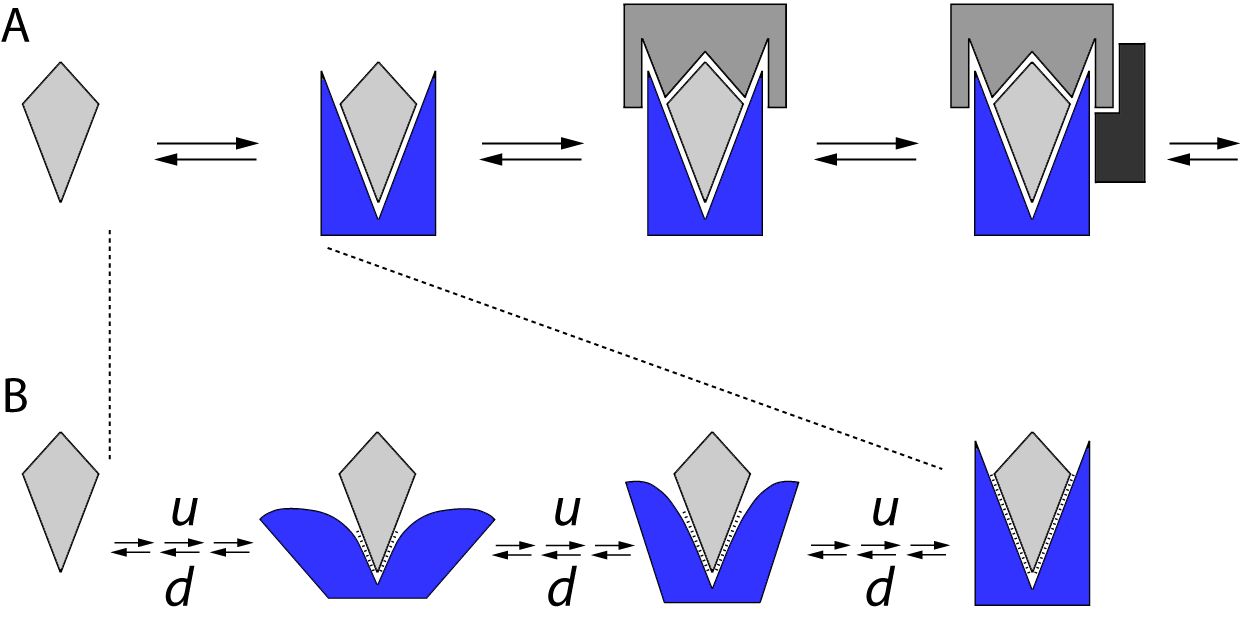} \\
\end{center}
\begin{small} \textbf{Figure 1.} Hierarchical multimolecular assembly with progressive adjustment and capping. (\textbf{A}) Components, or pre-made polymeric building blocks, are added one by one when the previous components are conveniently arranged. By this way, the nucleating complexes are trapped and prevented to dissociate. (\textbf{B}) Induced-fit interaction through progressive zippering of chemical bonds (rates $ u $ for upstream), counteracted by resistance to deformation (rates $ d $ for downstream). Only the most adjusted complex is capable of accommodating the next component (panel A), which in turn locks the preceding stepwise adjustment chain. \end{small}\\
\newline

\begin{center}
\includegraphics[width=6cm]{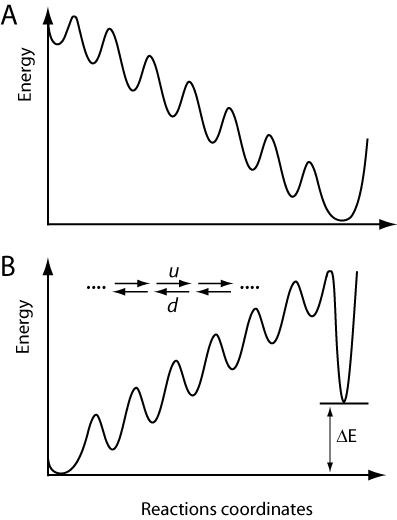} \\
\end{center}
\begin{small} \textbf{Figure 2.} Energy landscapes of: (\textbf{A}) exothermic vs (\textbf{B}) endothermic stepwise adjustment. The former, traditional case, is inherently stable and associated to a dissipation of heat measurable by calorimetry. By contrast, in the later case, a complex of high energy should be locked by a quasi-irreversible final step (deep right well). Note that in this scheme, the energy of the final step can be higher than the starting one, which means that such a system is capable of catching and storing thermal fluctuations. \end{small} \\
\newline

\subsection{Enzyme-substrate adjustment prior to reaction}
The mechanism described here differs from \cite{Hopfield,Ninio1975}, in that (i) there is a single entry of the substrate in the scheme, and (ii) many steps of conformational adjustment after initial binding. This adjustment is supposed to be necessary for the enzyme-substrate complex to react. The transformation reaction ensures the micro-irreversible (non-equilibrium) nature of the whole scheme, represented below with many successive states of the enzyme $ E $,

\begin{center}
\ce{\textit{E}_{0}
+\textit{S}
<=>[\ce{\textit{u}_{0}}][\ce{\textit{d}_{1}}]
$\ce{\textit{E}_{1}}$
<=>[\ce{\textit{u}_{1}}][\ce{\textit{d}_{2}}]
$\ce{\textit{E}_{2}...}$
<=>[\ce{\textit{u}_{n-2}}][\ce{\textit{d}_{n-1}}]
$\ce{\textit{E}_{n-1}}$
->[\ce{\textit{u}_{n-1}}]
$\ce{\textit{E}_{n}}$
+\textit{P} }
\end{center}
\begin{small} \textbf{Figure 3.} The multistep adjustment of the enzyme to its substrate is a prerequisite for the reaction to proceed. \end{small} \\
\newline
The enzyme state properly arranged for reacting is $ E_{n-1} $. $ E_{0} $ is waiting for a substrate, $ E_{1} $ is bound to a substrate, the intermediate states $ E_{1} $ to $ E_{n-1} $ are more and more adjusted enzyme-substrate complexes, of which only the last one can react to give the product. $ E_{n} $ has just released a product or incorporated the substrate. The forward ($ u $) and backward ($ d $) rates are numbered in reference to the starting states. $ u_{0} $ is a pseudo-first order constant including the concentration of the abundant substrate $ [S] $, whereas all the other constants $ u $ and $ d $ are genuine first-order kinetic constants.\\
As described below, remarkable biochemical behaviors are provided by these chains. In the steady state modeling used here, individual component concentrations result from constant turnovers of synthesis/degradation and the rate of degradation of the complexes is considered of the same order as their rate of synthesis. 

\begin{center}
\includegraphics[width=8cm]{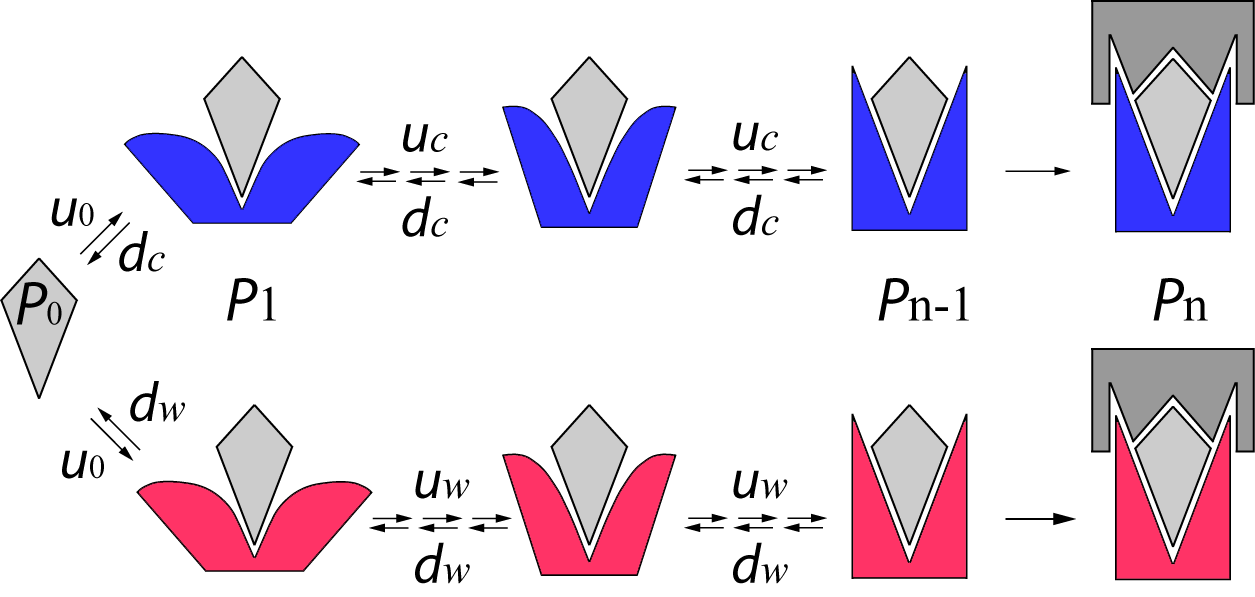} \\
\end{center}
\begin{small} \textbf{Figure 4.} Competition for binding between a correct ($ c $) and wrong ($ w $) partner supposed to be at equal concentration and uptaken from the medium with the same probability at rate $ u_{0} $. Every transition from $ P_{0} $ to $ P_{1} $ is the introduction of a candidate substrate; every transition from $ P_{1} $ to $ P_{0} $ is the rejection of the substrate, and the transition from $ P_{n-1} $ to $ P_{n} $ corresponds to the definitive incorporation of the substrate. This competition is simulated in the interaction tool "competitive binding" (Appendix A). \end{small} \\

\section{Substrate selection}
Giving a substrate successive possibilities of dissociation before biological action, is a fundamental principle of proofreading \cite{Hopfield,Ninio1975}. The mechanism of competition between binding partners presented here falls into this framework. In the two examples described above, the relative efficiency of incorporation of correct vs wrong substrates represented in Fig.4 can be evaluated by discrete approaches, such as through the mean number of substrate rejection before incorporation (Section 5), but it is classically evaluated in a continuous approach through the mean global conversion rate of substrates into products, corresponding to the reciprocal of the mean time of arrival $ \left \langle T \right \rangle $. $ \left \langle T \right \rangle $ can be calculated through random walk approaches, where the rates of forward ($ u $) vs backwards ($ d $) jumps are energetically related to each other through
\begin{equation} \dfrac{u_{i}}{d_{i+1}} = \textup{e}^{(E_{i}-E_{i+1})/k_{B}T} \end{equation}
\noindent
which says that the formation of a complex is favored when its energy is lower than that of the isolated components ($ u > d $). When $ u < d $, association reactions are less frequent but remain possible. They are facilitated by temperature increase at the macroscopic level and by fluctuations at the microscopic level. For the enzymatic reaction, in the stepwise enzymatic scheme, $ E_{0} $ and $ E_{n} $ are in fact indiscernible. Hence, the mean frequency of enzyme state recycling from $ E_{0} $ to $ E_{0} $ with release of one product $ P $, is the inverse of the mean time $ \left \langle T \right \rangle $ necessary to go from $ E_{0} $ to $ E_{n} $. The shortest is this time, the most probable is the transformation of the substrate into a product. We have shown in \cite{MichelR} that $ \left \langle T \right \rangle $ obeys the very general formula

\begin{equation} \left \langle T \right \rangle=\sum_{i=0}^{n-1}\sum_{j=0}^{n-i-1}\frac{1}{d_{j}}\prod_{k=j}^{i+j}\frac{d_{k}}{u_{k}} 
\label{meanT}
\end{equation}
which naturally depends on the ratio between the upstream and downstream rates. In particular, $ \left \langle T \right \rangle $ dramatically increases when backward transitions are slightly more probable than forward ones. To easily evaluate the influence of the ratio $ d/u $ on $ \left \langle T \right \rangle $, Eq.(3) can be advantageously rewritten in an elegant form reminiscent of the pioneering probabilistic approaches of \cite{Ninio1975,Knorre} 

\end{multicols}
\begin{equation} \left \langle T \right \rangle =\sum_{i=0}^{n-1}\dfrac{1}{u_{i}}\left (1+\dfrac{d_{i+1}}{u_{i+1}}\left (1+\dfrac{d_{i+2}}{u_{i+2}}\left (.... \left (1+\dfrac{d_{n-1}}{u_{n-1}}  \right )....\right )  \right )  \right )  \end{equation}
\begin{multicols}{2}
\noindent
In most real situations, all the individual reaction rates are to some extent different, but in the case of induced-fit, the individual micro-reactions are single noncovalent chemical bond formations, which can be approximated as roughly equivalent. Hence, we will consider that all the forward rates are identical ($ u $) and all the backward rates are identical ($ d $). The kinetics of achievement of this simplified chain can be calculated using Laplace transforms \cite{Munsky,Bel}, direct matrix (Appendix B), or more simply recovered by reduction of the powerful Eq.(4), as shown below.

\subsection{Easy binding through predominantly forward walk}
When the backward transitions are quasi-inexistent ($ d_{i} \simeq  0 $), the random walk is radically forward and completed very rapidly and Eq.(4) simplifies as

\begin{subequations} \label{E:gp}  
\begin{equation} \left \langle T \right \rangle =\sum_{i=0}^{n-1}\dfrac{1}{u_{i}} \end{equation} \label{E:gp1}
\noindent
For identical $ u_{i} $, 
\begin{equation} \left \langle T \right \rangle = \frac{n}{u} \end{equation} \label{E:gp2}
\end{subequations} 

The mean time of achievement of the chain is simply the sum of the waiting times of the individual steps.

\subsection{Symmetric random walk}
In the particular case where all the reverse and forward rates are all identical, ($ d/u=1 $), Eq.(4) reduces to

\begin{subequations} \label{E:gp}
\begin{equation}  \left \langle T \right \rangle =\dfrac{1}{u}(\underbrace{[1]+[1+1]+[1+(1+1)]+...}_{n \ \textup{pairs of brackets}}) \end{equation} \label{E:gp1}
\noindent
That is, given the sum of the $ n $ first integers,
\begin{equation}  \left \langle T \right \rangle = \dfrac{1}{u}\frac{n(n+1)}{2} \end{equation} \label{E:gp2}
\end{subequations} 

In the two cases examined above ($ d=0 $ and $ d/u=1 $), substrate binding depends only on $ u_{0} $, so that the possible differences between competing molecules cannot be further discriminated. A much more interesting situation is obtained for a strong tendency to dissociate. 

\subsection{Endothermic adjustment}
If there is a strong resistance to adjustment ($ d \gg u $), Eq.(4) simply becomes

\begin{subequations} \label{E:gp}  
\begin{equation} \left \langle T \right \rangle \sim \dfrac{1}{u}\sum_{i=0}^{n-1} \left (\frac{d}{u} \right )^{i} \end{equation} \label{E:gp1}
\noindent
that is dominated for large $ d/u $, by the larger component
\begin{equation} \left \langle T \right \rangle \sim \dfrac{1}{u} \left (\frac{d}{u}  \right )^{n-1} \end{equation} \label{E:gp2}
\end{subequations} 

Strikingly in this type of random walk, the walker is most often found at the first step or the origin of the walk, which means that the binding partner is often rejected. The walk can occasionally progress until the $ n $th step in a probabilistic manner. Eq.(7b) is a geometric function of the number of steps, which means that small differences in conformational accommodation, are strongly amplified. By this way, self-assembly occurs difficultly but properly whereas the incorporation of wrong components is precluded. The simulations presented in Appendix A clearly illustrate the critical importance of the balance between backward and forward transition rates for achieving this goal.

\section{Back to zero} 
Another way to compare the capacity of substrate selection of the different walks is to define the function linking the ratio $ d/u $ to the mean number of substrate rejections before fixation. In case of competition between a correct and wrong substrate (Fig.4), the probability of rejection directly reflects the average number of returns to stage 0 before the first arrival to stage $ n $. Each rejection of the wrong substrate is a new opportunity to bind the correct substrate. It is therefore important that the mean number of rejections of the bad substrate is much higher than the number of rejections of the good one. This number can be obtained through an original discrete method described in Appendix C. It shows that the representation ratio between any two states $ i $ and $ j $ preceding incorporation, is

\begin{subequations} \label{E:gp}  
\begin{equation} \dfrac{\left \langle N_{i} \right \rangle}{\left \langle N_{j} \right \rangle} = \dfrac{1-\left (\frac{d}{u} \right )^{n-i}}{1-\left (\frac{d}{u} \right )^{n-j}}\end{equation} \label{E:gp1}
\noindent
The rarest state in the chain is the state $ n-1 $ and the number of substrate rejections before incorporation (state $ 0 $), is higher than the number of states $ n-1 $, according to the ratio
\begin{equation} \dfrac{\left \langle N_{0} \right \rangle}{\left \langle N_{n-1} \right \rangle} = \dfrac{1-\left (\frac{d}{u}  \right )^{n}}{1-\frac{d}{u}} \sim \left (\frac{d}{u}  \right )^{n-1} \end{equation} \label{E:gp2}
\end{subequations} 

For $ d/u = 1 $, this ratio tends to $ n $, but it dramatically increases for $ d/u $ exceeding unity. Fig.5 shows the relative number of passages in the different states of a short walk of only 6 steps ($ i=0,1,2,3,4,5 $). Hence, repetitive returns to the starting point become astonishingly numerous in the predominantly backward random walks, like for the mythic Sisyphus, who had to push up a rock rolling again and again down the mountain. The same scenario holds for the thermodynamic mountain of the backward walk, except that since there are many Sisyphus, some of them can occasionally succeed in reaching the top of the mountain. 

\begin{center}
\includegraphics[width=7cm]{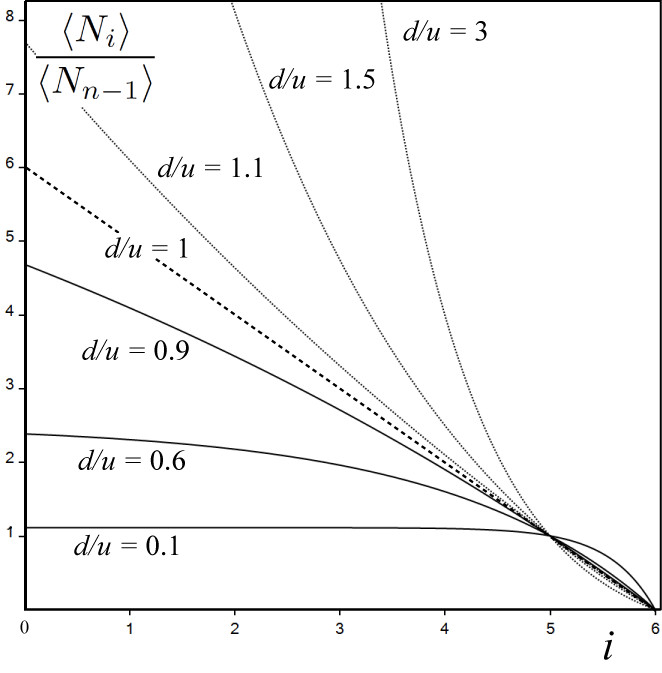} \\
\end{center}
\begin{small} \textbf{Figure 5.} Number of passages in state $ i $ compared to $ n-1 $, the less represented state in a finite homogeneous walk, for different values of $ d/u $. In the example shown, $ n=6 $.  When $ d>u $, this mechanism allows to strongly separate two resembling substrates with slightly different affinities. \end{small} \\
\newline

\section{Time dispersion of the arrivals}
In addition to modify the mean completion time and the mean number of rejections, the ratio $ d/u $ has also a great impact on the dispersion of the arrivals, with interesting biological consequences. This parameter, illustrated by the standard deviation, is summarized in Table 1.\\

\subsection{Predominantly forward walk}
The strictly forward, no-return walk has the remarkable property of being the most focused in time, compared to a single transition of equivalent waiting time. This property is an advantage in certain circumstances such as signal transduction \cite{Doan}. But it is likely to be not desirable for the building of complexes because: (i) the complexes would be often defective following misincorporation of incorrect components and (ii) their synthesis would be synchronous and massive, thereby hardly manageable by the cell. The "randomness parameter" $ r $ defined as $ (\sigma (T)/\left \langle T \right \rangle)^{2} $ \cite{Floyd}, is $ 1/n $ for forward random walks, but it is independent of the number of steps for symmetric and backward walks.

\subsection{Symmetric walk}
The standard deviation varies as the square of the number of steps, according to the long established property of diffusion.

\subsection{The predominantly backward walks}
For a predominantly backward walk, the standard deviation is a geometric function of the number of steps, just like the mean time of completion (Table.1), demonstrated in Appendix B. This result means that individual arrivals are widely scattered in time. As a consequence, multimolecular complexes are expected to appear sporadically in the different areas of the cell where its components are diffusing, thus preventing sudden particle accumulation after increased synthesis of their constituents. This dispersion, associated to a convenient lifetime of the complexes, also prevents the complete removal of free components from the medium.\\

\begin{small}\textbf{Table 1.} Mean time of arrival and its standard deviation for backward, symmetric and forward finite random walks.
\end{small}
\begin{center}
\medskip
\begin{tabular}{lll}
\hline\noalign{\smallskip}
Walk & $ \left \langle T \right \rangle $ & $ \sigma (T) $  \\
\noalign{\smallskip}\hline\noalign{\smallskip}
Predom. forward ($ u \gg d $) & $  \frac{1}{u} n $ & $ \frac{1}{u} \sqrt {n} $ \\
Symmetric ($u \simeq d$) & $ \frac{1}{u} \frac{n(n+1)}{2} $ & $ \frac{1}{u} \frac{n^{2}}{\sqrt {6}} $ \\
Predom. backward ($ u \ll d $) & $ \frac{1}{u} \left (\frac{d}{u}  \right ) ^{n-1} $ & $   \frac{1}{u}  \left (\frac{d}{u}  \right )^{n-1} $ \\
\noalign{\smallskip}\hline
\end{tabular}
\end{center}
\noindent

\section{Near absence of incomplete complexes in the cell}
In addition to their differential capacity of buffering bursts of complex synthesis, the predominantly forward and backward chains differ in the representation of partial complexes. For a backward random walk, the arrivals of new complexes in the cell are spread in time, but the appearance of a single particle is both rapid and complete, thus avoiding the presence in the cell of possibly harmful or dominant negative complexes under construction. In this scenario, the components expected to prevail in the cell are either very small or complete multimers, but partial complexes lacking a few components are very transient. A simulation applet named "Intermediate states" allows to visualize the dramatic influence of the ratio $ d/u $ on the different state probabilities along the chain (Appendix A, Fig.A2).

\section{Thermodynamic cost of finite backward random walks}
As shown above, the backward random walk presents many advantages for biochemical systems. It is a nonequilibrium process due to the irreversible final step, but it does not necessitate NTP-consuming active mechanisms. If binding selectivity can be obtained for free, one may wonder why the first proposed mechanisms of substrate selection are energetically expensive, concretely illustrated by NTP consumption \cite{Hopfield,Blomberg}. A possible explanation is that energy consumption concerns polymerases (ribosomes or RNA polymerases), whose polymerization rates (of translation or transcription), can not be too low. Their coupling to energy-providing reactions allows to rapidly check the quality of subtrates without waiting for spontaneous positive fluctuations. By this way, the delay of synthesis of RNAs or proteins remains reasonable enough not to affect the reactivity of gene regulatory circuits. By contrast, there are no temporal constraints on the rates of complex formation which can be arbitrary, provided they are associated with appropriate rates of removal to maintain acceptable stationary concentrations of complexes in the cell. In this case, the energy necessary for endothermic reactions can be uptaken from spontaneous thermal fluctuations. At nonzero temperature, thermal and density fluctuations are the rule; but at equilibrium they are reversible. Hence, the essential aspect of the principle described here is the capture of fluctuations by final micro-irreversible steps acting as ratchets, already identified as essential biochemical mechanisms in the field of molecular motors \cite{Herzog}. The backward random walk mechanism has nevertheless a cost, but which has in fact already been paid in the past during the evolutionary design of lock-and-key ratchet components, through information-retrieval cycles \cite{Michel_Brillouin}. 

\section{Extension of the principle of endothermic chains to other situations}

The substrate selection mechanism described here relies on the delay of the final step of the chain. It offers the opportunity to multiply substrate "re-weighting" processes to distinguish competing interactions \cite{Ninio1975}. In addition, long backward micro-irreversible chains can ensure a multitude of other essential biological roles. Examples of intracellular processes to which it applies are listed below.

\subsection{Preparing complexes subject to rapid depolymerization}

Certain complexes are made of a large number of the same building unit. The concentration of this unit directly sets the value of the upstream pseudo-first order rate $ u $ which is the same all along the polymerization chain. A clear example is the formation of microtubules. Over a certain concentration of elementary microtubule building blocks ($ \alpha \beta $ tubulin dimers), the formation of microtubular fibrils proceeds in a seemingly stochastic manner, just like the backward random walk. More energy is clearly stored in the microtubules compared to diffusing tubulin. This energy has been shown to derive from the hydrolysis of GTP \cite{Caplow}, but interestingly, microtubules can also form in absence of GTP hydrolysis and in this case it has been shown endothermic, as determined by microcalorimetry \cite{Hinz}. The initial steps of microtubule formation are considered as counter-thermodynamic \cite{Flyvbjerg}. Microtubules are then stabilized by accessory apical proteins locking the microtubule ends. Remarkably, the destabilization of microtules, partly prevented by GTP-containing subunits at the growing end of microtubules, leads to explosive depolymerizations named catastrophes. Non-hydrolysed GTP remnants along the microtubule have been proposed to prevent complete depolymerization \cite{Dimitrov} and to reinitiate rescues. Rapid microtubule shrinking plays important biological roles \cite{Coue}, consistent with the fact that tubulin dissociation is exothermic. Remarkably, polymerization of pure tubulin can be spontaneous \cite{Hinz} by catching the energy of thermal fluctuations and in turn, the energy stored in microtubules is exploited during rapid cellular reorganizations.

\subsection{An additional mechanism involved in protein folding}

Protein binding and protein folding are closely related processes. According to current models, the initial stages of soluble protein folding are hydrophobic collapses, which are followed by the arrangement of surrounding domains. The progressive establishment of these inter-domain interactions can proceed through stepwise adjustement, as for the association of distinct proteins. 

\subsection{Kinetic genesis of isogenic heterogenetity}
The dramatic change of standard deviation obtained when shifting from forward to backward random walk (bottom last line of Table.1), has a great interest in the field of isogenic heterogeneity. This point is illustrated by the induction of the lactose operon (\textit{lac}). The molecular event triggering \textit{lac} induction is the complete dissociation of the transcriptional repressor (LacI) from the \textit{lac} DNA. As LacI contains four monomers which all contact \textit{lac}, complete LacI dissociation necessitates the achievement of a chain of LacI monomer dissociation. Complete dissociation can then be maintained by a positive feed back and is necessary for full and durable transcriptional induction and to the switch of certain bacteria into a self-sustained induced state \cite{Xie,Michel_lactose}. This chain of dissociation events is controlled by antagonistic tendencies: (i) of dissociation provoked by the inducer (modifying the pseudo-first order rates $ u $) and (ii) of strong association between LacI and \textit{lac} \cite{Michel_lactose}. Hence, at high inducer concentration, all the bacteria are expected to switch synchronously to the induced state, whereas at low inducer concentration, there is a strong resistance to LacI dissociation, so that only certain bacteria become induced in an unpredictable manner. Moreover, owing to the positive feedback of the lactose operon, these few activated bacteria remain fully and stably induced \cite{Xie,Michel_lactose}. This purely kinetic mechanism has a strategic importance at the population level as it can contribute to the bimodality of the lactose operon induction at low doses of inducer, without need for the conventional mode of bistability \cite{Michel_lactose}, like other examples of extended transient bimodality in monostable systems \cite{Tiwari}.

\section{Origin and role of the micro-irreversible step}

The micro-irreversible final steps of the chains described here act as thermodynamic ratchets. Their role is essential for catching fluctuations as they make the difference between inexistent Maxwell demons and realistic biochemical mechanisms \cite{Michel_Brillouin}. Their origins are multiple, as illustrated by the different examples used here. 
\begin{itemize}\setlength{\itemsep}{0.5mm}
  \item \textit{For hierarchical complex building}. Micro-irreversibility is ensured by locking the most adjusted complex (Fig.1). This capping phenomenon allows to (i) initiate the next step and (ii) freeze previous interactions, even if they are not very stable by their own. Capping steps are likely to be scattered over the assembly line of large molecular complexes.
  \item \textit{For enzymatic reactions}. Enzymatic reactions have long been modeled in (Fig.3) as micro-irreversible. The irreversibility of the transition $ u_{n-1} $ is not related to some irreversibility in the catalytic reaction itself, but to the escape of the product $ P $ once formed. The very low concentration of $ P $ in the cell prevents it to rebind to the enzyme. Indeed, $ P $ is generally immediately used as a substrate for a subsequent reaction, or incorporated into larger macromolecules.
  \item \textit{For tubulin polymerization} Storing thermal fluctuations is rendered possible by ratchet mechanisms preventing their immediate dissipation. GTP-containing subunits and various tip proteins have been proposed to ensure this role for microtubules.
  \item \textit{For the example of the lactose operon derepression} (Section 9.3), the very low concentration of free LacI in the cell strongly delays its rebinding to \textit{lac} after dissociation \cite{Xie}. This long delay can allow sufficient expression of the operon to definitely prevent LacI-\textit{lac} rebinding through the famous positive feedback locking \textit{lac} in the induced state \cite{Horibata,Xie,Michel_lactose}.
\end{itemize}
The micro-irreversibility of the first arrival to the final stage of the chain, works as a fluctuation ratchet stabilizing a low probability state. By this way, counter-thermodynamic reactions can be achieved without need for the conventional recipe of biochemistry of a coupling with the consumption of an energy donor such as ATP or GTP.

\section{Discussion}
The hallmark of life is its capacity to locally decrease entropy, or in other words, to select improbable states. This is precisely what is achieved here when states of low probability are functionally selected. Inappropriate interactions are generally assumed to be excluded in lock-and-key interactions by the much higher affinity for relevant binding partners but this principle suffers from several drawbacks: (\textbf{i}) The requirement of a perfect stereo-complementarity is not suitable for early, poorly evolved living systems in the context of the origin of life. (\textbf{ii}) Partner discrimination in the lock-and-key mechanism, is possible only at the initial interaction (rates $ k_{\textup{on}} $ and $ k_{\textup{off}} $) which are clearly insufficient to distinguish between resembling substrates \cite{Hopfield,Ninio1975}. (\textbf{iii}) As a matter of fact, many closely related proteins coexist in the cell, for example deriving from multigenic families, and their discrimination requires amplifying faint differences of dissociation rates. Difficult induced-fit is well suited for this purpose, as it allows discriminating competing binding partners, not only through the initial association, but more importantly on following conformational adjustment that is spread over many sticking micro-reactions. 
The gain of discrimination thus obtained is not merely additive, but multiplicative, as clear in Eq.(7b) and in accordance with the illuminating analogy of the coin reweighing machines proposed in \cite{Ninio1986} for detecting false coins. This recipe relies on the fundamental property of probabilities, which also underlies entropy: the probability of a union of independent events is the products of the probabilities of the individual events. This mode of discrimination resembles the proofreading mechanisms formerly described \cite{Hopfield,Ninio1975} in that it delays the final reaction of molecular incorporation. But beside this common feature, it differs through other points: (\textbf{i}) Contrary to \cite{Hopfield}, no "hard-driven" energy-consuming step is required. (\textbf{ii}) The number of "discrimination steps" is not limited as in the mechanisms of \cite{Hopfield,Ninio1975}, but covers a very large series of elementary transitions. \\
The predominantly backward finite random walk predicts a probabilistic mode of assembly with several features: (\textbf{i}) The building of supramolecular structures resembles an all-or-none phenomenon, which presents the advantage of reducing the abundance of incomplete complexes in the cell. (\textbf{ii}) Complete particles are expected to appear in a sporadic manner in the cell, avoiding jolts in their synthesis. (\textbf{iii}) In this model of energy storage, the core of supramolecular complexes could continuously "breath" through antagonistic constraints. Such "living" complexes could have interesting properties absent from rigid lock-and-key complexes. The so-called dynamic instability of microtubules is fundamentally related to the energy content of this complex.\\ 
One of the essential virtues of the mechanism proposed here is to allow discriminating resembling interaction partners more surely than a single lock-and-key binding. This principle is likely to be a general recipe allowing living systems to filter inappropriate interactions.\\
\newline

\end{multicols}

\newpage

\begin{center}
\Huge{Appendices}
\end{center}

\appendix

\section{Discrete simulation}

To clearly appreciate the crucial role of small differences between backwards and forward transitions in the final achievement of finite random walks, A series of interactive simulation tools in discrete time is developed, which agrees with the mathematical predictions in continuous time and allows to clearly visualize the singular behavior of predominantly backward finite random walks. The set of applets using the simulation platform SimuLab (Observatoire de Paris, http://media4.obspm.fr/outils/simulab/) and accessible in: http://selfassembly.genouest.org/. The tool "Complex formation" (Fig.A1A) allows to compare the accumulation in space of correct vs wrong complexes, while the simulation "Competitive binding" (Fig.A1B) shows the comparative kinetics of incorporation of wrong and correct substrates and their final incorporation levels.

\begin{center}
\includegraphics[width=15cm]{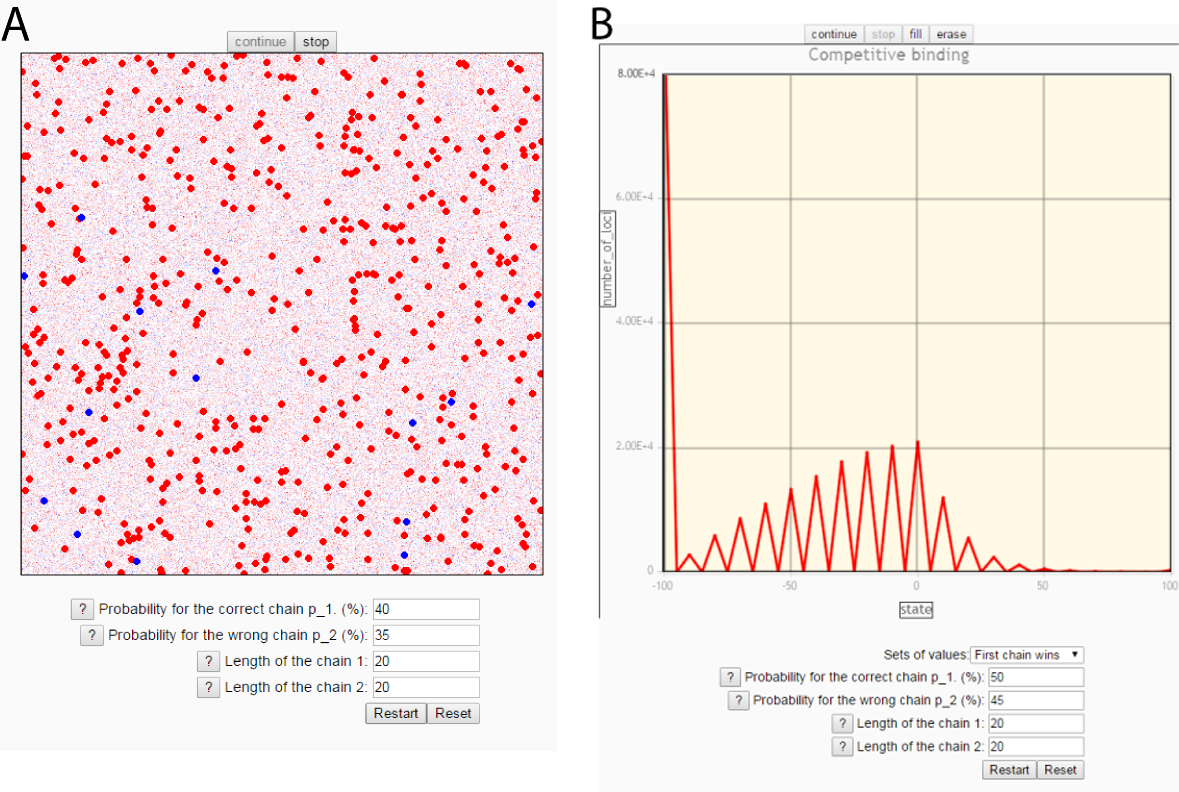} \\
\end{center}
\begin{small} \textbf{Figure A1.} Stochastic simulation of binding selectivity (\textbf{A}) Screen capture of the tool named Complex formation, showing the great advantage conferred by small differences between backward and forward transitions. In spite of the slight difference between the ratios $ u/d $, the incorrect substrate is rarely incorporated (blue, darker spots). (\textbf{B}) Snapshot of the tool named Competitive binding, showing a competitive substrate incorporation. The correct and wrong substrates are mixed in the center of the graph and their differential capacity of incorporation in the final state is compared in left and right parts of the graph respectively . \end{small} \\
\newline
 Single chain evolution tools are also proposed, among which the tool "Intermediate states" allows to appreciate the relative representations of the different states of complex association before final locking (Fig.A2) and the tool "Total number of 0" shows the evolution of the number of substrate rejections before incorporation, obtained for different ratios $ d/u $ (Fig.A3).

\begin{center}
\includegraphics[width=7cm]{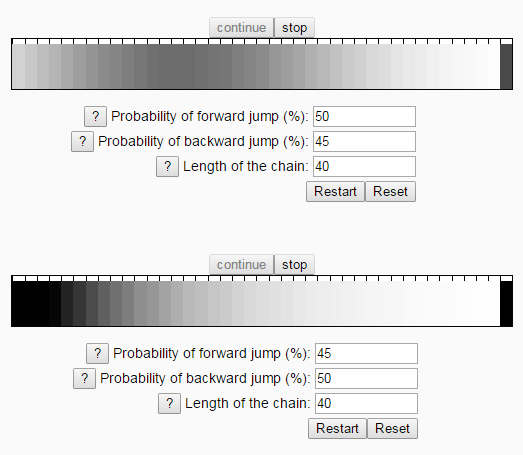} \\
\end{center}
\begin{small} \textbf{Figure A2.} Snapshot of the degree of occupancy of each state during a 40 step-walk, compared for a predominantly forward (top, $ u/d =0.50/0.45 $) and predominantly backward walk (bottom, $ u/d =0.45/0.50 $) at the mid time point of achievement. The starting state is at left and the final state is at right. Large intermediate states are extremely rare for the backward walk. In the forward walk, a wave of occupancy progresses through all the states. This occupancy tends to equalize between the different states when $ d<<u $. In biochemical terms, this equal occupancy means for example that all the sizes of a polymer would coexist in the cell. \end{small} \\
\newpage

\section{Mean completion times and standard deviations in the homogeneous case}

The general equations (3) and (4) from which the mean times of arrival $\langle T \rangle$ have been deduced in this article, have been demonstrated in (Michel and Ruelle, 2013). They are of course simpler in the homogeneous case for which all forward and backward rates are equal, $u_i=u$ and $d_i=d$. The standard deviation related to the first and second moments are derived below in this simplified case. Note that they can be equivalently obtained using Laplace transforms (Bel et al. 2009; Munsky et al., 2009). The matrix $A$ describing the vector of the differential system $\vec P(t) = (P_0(t),P_1(t),\ldots,P_{n-1}(t))$ as
\be
{{\rm d} \over {\rm d}t} \vec P(t) =  A \vec P(t)\,, 
\ee
simplifies in the homogeneous case and essentially depends on the ratio $\kappa \equiv {d \over u}$,
\small
\be
A = u \left(\begin{matrix}
-1 & \kappa & 0 & \ldots & 0\cr
1 & -(1+\kappa) & \kappa & \ldots & 0\cr
0 & 1 & -(1+\kappa) & \ldots & 0\cr 
0 & 0 & 1 & \ldots & \ldots \cr
\ldots & \ldots & \ldots & \ldots & \kappa \cr
0 & 0 & 0 & \ldots & -(1+\kappa)\cr
\end{matrix} \right).
\ee

\normalsize
This matrix, tridiagonal and almost constant along the main three diagonals, is simple enough to be explicitly inverted $B \equiv A^{-1}$. One finds
\be
B_{i,j} = {-1 \over u} \cdot {\kappa^{\max(0,j-i)} - \kappa^{n-i} \over 1-\kappa}, \quad 0 \leq i,j \leq n-1.
\label{Ainv}
\ee
The last row and first column of $B$ read
\bea
&& \hspace{-1cm} B_{n-1,i} = -{1 \over u},\\
&& \hspace{-1cm} B_{i,0} = -{1 - \kappa^{n-i} \over u(1 - \kappa)} = -{1 \over u} \, (1 + \kappa + \ldots + \kappa^{n-1-i}),
\eea
from which we obtain the mean time of arrival,
\be
\langle T \rangle = {1 \over u} \sum_{i=0}^{n-1} \:(1 + \kappa + \ldots + \kappa^{n-1-i}) = {n - (n+1)\kappa + \kappa^{n+1} \over u\, (1-\kappa)^2}.
\ee
When the walk is strongly forward ($\kappa \ll 1$), close to being symmetric ($\kappa \sim 1$) or strongly backward ($\kappa \gg 1$), the behavior of the mean arrival time is
\be
\langle T \rangle = \begin{cases}
{n \over u} & \mbox{for } \kappa \ll 1, \\
\noalign{\smallskip}
{n(n+1) \over 2u} & \mbox{for } \kappa \sim 1, \\
\noalign{\smallskip}
{1 \over u}\kappa^{n-1} & \mbox{for } \kappa \gg 1.
\end{cases}
\label{Thomog}
\ee

The calculation of the second moment $\langle T^2 \rangle$ is slightly longer but similar. In the general homogeneous case, we find
\be
\langle T^2 \rangle = {2 \over u^2(1-\kappa)^3}\left\{ {n(n+1)\over 2}(1-\kappa) +  3n\kappa^{n+1} - \kappa(2+\kappa^{n+1}) {1-\kappa^n \over 1-\kappa}\right\}.
\ee
In the three asymptotic regimes, it behaves as (the dominant terms given below do not assume $n$ large)
\be
\langle T^2 \rangle = \begin{cases}
{n(n+1) \over u^2} & \mbox{for } \kappa \ll 1, \\
\noalign{\smallskip}
{n(n+1)(5n^2+5n+2) \over 12u^2} & \mbox{for } \kappa \simeq 1, \\
\noalign{\smallskip}
{2 \over u^2} \kappa^{2n-2} & \mbox{for } \kappa \gg 1.
\end{cases}
\label{T2homog}
\ee
The value of the standard deviation $ \sigma(T) = \sqrt{\langle T^2 \rangle - \langle T \rangle^2}$ follows from (\ref{Thomog}) and (\ref{T2homog}) and, for $n$ large, given in Table 1.
\newpage

\section{Relative representation of intermediate states along the walk} The chain of reactions considered in the text can be viewed as the continuum limit of a discrete time random walk, and can therefore be given a probabilistic interpretation. In order to see this in concrete terms, we define the stochastic evolution of a hopping particle. The particle may be found at one of $n+1$ sites (or states), numbered from 0 to $n$, and jumps from site to site according to specific random rules. The particle makes one jump every time interval, moving from site $i$ to site $j$ with a certain probability $p_{ji}$. In the present context, the hopping particle, also referred to in the literature as a walker, is just a way to see a certain amount of product $i$ change to product $j$, the change occurring with probability or rate $p_{ji}$ (in the limit of continuous time, the probabilities will become instantaneous transition rates). Equivalently, we can say that the fraction $p_{ji}$ of product $i$ transforms to product $j$. 

Let us now define more precisely the jumping rules, using the walk terminology, more convenient and intuitive. The walker is initially at site ${i_0}$ at time $N=0$, and from then on, takes one step every time interval. However the only steps he is allowed to take are $\pm 1$ or $0$, with jumping probabilities
\be
p_{ji} = p(i \to j) = \begin{cases}
\upsilon _i & \hbox{if } j=i+1,\\
\delta_i & \hbox{if } j=i-1,\\
1-\upsilon _i-\delta_i & \hbox{if } j=i.
\end{cases}
\ee
If the walker position at time $N$ is $i$, his position at the next time $N+1$ can only be $i-1,i,i+1$. Moreover the associated probabilities are in general different for the three steps (and site-dependent). Because the positions should remain in the set $\{0,1,\ldots,n\}$, we must specify the boundary conditions at the two end-points of the chain. We decide that site $i=0$ is reflecting by setting $\delta_0=0$, and that site $i=n$ is absorbing by choosing $\delta_n=\upsilon _n=0$, meaning that once the walker reaches site $n$, he stays there forever, with probability 1. 

The position of the walker at any fixed time $N$ is a random variable, characterized by a distribution ${\cal P}_{i_0}(\bullet;N)$, where ${\cal P}_{i_0}(i;N)$ denotes the probability that the walker be at site $i$ at time $N$, having started from site ${i_0}$. The initial distribution is ${\cal P}_{i_0}(i;0) = \delta_{i,{i_0}}$, and each fixed time distribution is normalized, $\sum_{i=0}^n {\cal P}_{i_0}(i;N) = 1$ for any ${i_0},N$.

Simple probabilistic arguments show that the distributions satisfy the following discrete time evolution equation,
\be
{\cal P}_{i_0}(i;N+1) = \upsilon _{i-1} \, {\cal P}_{i_0}(i-1;N) + \delta_{i+1} \, {\cal P}_{i_0}(i+1;N) + (1-\delta_i-\upsilon _i) \, {\cal P}_{i_0}(i;N).
\label{evol}
\ee
With the initial condition given above, it can be proved that the distributions at later times are uniquely determined, and can be computed by standard methods. We observe that the previous recurrence equation can be written in a matrix form as
\be
{\cal P}_{i_0}(i;N+1) = \sum_k ({\mathbb I} + \hat{\cal A})_{i,k} \, {\cal P}_{i_0}(k;N),
\ee
where the entries of ${\mathbb I} + \hat{\cal A}$ are precisely the transition probabilities, $({\mathbb I} + \hat{\cal A})_{i,k} = p_{ik}$. By iterating the recurrence, we obtain a formal but useful expression, 
\be
{\cal P}_{i_0}(\bullet;N) = ({\mathbb I} + \hat{\cal A})^N \, {\cal P}_{i_0}(\bullet;0),
\label{solu}
\ee
written in vector notation (${\cal P}_{i_0}(\bullet;N)$ is a column vector with components ${\cal P}_{i_0}(i;N)$).

Though it may be tricky to obtain explicit expressions when $\delta_i,\upsilon _i$ are arbitrary, the gross features of the behavior for large times may be established by elementary means. In the generic case, namely when the probabilities $\upsilon _i$ are all non-zero, we note that the walker has a non-zero probability $p_a$ to be absorbed in time $n$ (whatever its starting point), and therefore a probability $p_s=1-p_a$ strictly smaller than 1 to survive up to time $n$. By considering temporal windows which are multiples of $n$, the survival probability up to time $kn$ ($k$ integer) is bounded by $p_s^k=(1-p_a)^k$. It follows that the survival probability up to time $N$ decays exponentially with $N$. Sooner or later, the walker will be absorbed, with probability 1, or $\lim_{N \to \infty} {\cal P}_{i_0}(i;N) = \delta_{i,n}$.

Our purpose here is two-fold. First we show that the continuous time limit of the discrete random walk defined above precisely yields the deterministic reaction processes considered in the text. In a second step, we will be interested in a quantity that can be conveniently computed in the discrete formalism, namely the average number of returns to site 0 (or to any other site) before the walker gets eventually absorbed at site $n$. 

In order to define the continuous time limit of the random walk, we take the time increment to be a small quantity $\epsilon$, instead of 1 as was done above, and let the walker take a step every $\epsilon$ unit of time. The time elapsed after $N$ moves is $t=N\epsilon$. Eq. (\ref{evol}) then reads
\be
{\cal P}_{i_0}(i;\tfrac{t+\epsilon}{\epsilon}) - {\cal P}_{i_0}(i;\tfrac{t}{\epsilon}) = \upsilon _{i-1} \, {\cal P}_{i_0}(i-1;\tfrac{t}{\epsilon}) + \delta_{i+1} \, {\cal P}_{i_0}(i+1;\tfrac{t}{\epsilon}) - (\delta_i+\upsilon _i) \, {\cal P}_{i_0}(i;\tfrac{t}{\epsilon}).
\ee
For small and finite $\epsilon$, the time variable $t$ is discrete but comes closer and closer to a continuous variable as $\epsilon$ tends to 0. Dividing the previous equation by $\epsilon$ and taking the limit $\epsilon \to 0$ yields the following differential system
\be
{{\rm d} \over {\rm d}t}\,P_{i_0}(i;t) = u_{i-1} \, P_{i_0}(i-1;t) + d_{i+1} \, P_{i_0}(i+1;t) - (u_i+d_i) \, P_{i_0}(i;t),
\ee
where the discrete and continuous rates are related by $\upsilon _i = \epsilon\,u_i,\, \delta_i = \epsilon\, d_i$, and the continuous time distributions are defined from
\be
P_{i_0}(i;t) = \lim_{\epsilon \to 0} {\cal P}_{i_0}(i;\tfrac{t}{\epsilon}).
\ee
Let us note that the limit over $\epsilon$ in the previous equation is not simply a large $N$ limit of ${\cal P}_{i_0}(i;N)$ but also applies to the implicit dependence (not shown) of ${\cal P}_{i_0}(i;N)$ on the parameters $\delta_i,\upsilon _i$, so that the limiting distributions $P_{i_0}(i;t)$ depend on $d_i,u_i$.

For ${i_0}=0$, we recover the differential equations satisfied by the concentrations $P_i(t) \equiv P_0(i;t)$ recalled in (9) and (10). The relationship between the probability distributions and the concentrations considered in the context of the chain of reactions stems from the frequentist interpretation of probabilities. $P_i(t)$ may be viewed as the probability that a single complex is in state $i$ or in case of a large number of complexes, as the proportion of those in state $i$. 

Let us now compute ${\cal T}_{i_0}(i)$, defined as the average number of times the walker comes to site $i$ before being eventually absorbed (the starting site is $i_0$ while the time of absorption is not fixed). A simple and classical argument is sufficient to obtain a convenient formula. 

For each discrete time $N$, let us define the binary random variable $I(i;N)$: it is equal to 1 or 0 according to whether the walker is or is not at site $i$ at time $N$. Its distribution is simply Prob$[I(i;N)=1] = {\cal P}_{i_0}(i;N)$, implying that its average value is $\langle I(i;N) \rangle = {\cal P}_{i_0}(i;N)$. Because the total time spent by the walker at $i$ is equal to $\sum_{N=0}^\infty I(i;N)$ (if $i=i_0$), the initial time is included), the corresponding average value is 
\be
{\cal T}_{i_0}(i) = \sum_{N=0}^\infty \: \langle I(i;N) \rangle = \sum_{N=0}^\infty \: {\cal P}_{i_0}(i;N)= \sum_{k=0}^n \: \sum_{N=0}^\infty \: ({\mathbb I} + \hat{\cal A})^N_{i,k} \, {\cal P}_{i_0}(k;0),
\ee
where we have used (\ref{solu}) in the last step. 

To carry out the two summations, we may assume $i_0,i \neq n$ since otherwise ${\cal T}_{i_0}(i)$ is infinite. The summation over $k$ can then be restricted to $k<n$, and likewise the matrix $({\mathbb I} + \hat{\cal A})$ can be restricted to its $n \times n$ submatrix labeled by the sites $0,1,\ldots,n-1$, which we denote by $({\mathbb I} + {\cal A})$. We obtain
\be
{\cal T}_{i_0}(i) = \sum_{k=0}^{n-1} \: \sum_{N=0}^\infty \: ({\mathbb I} + {\cal A})^N_{i,k} \, {\cal P}_{i_0}(k;0) = - \sum_{k=0}^{n-1} \: ({\cal A}^{-1})_{i,k} \, {\cal P}_{i_0}(k;0) = - ({\cal A}^{-1})_{i,i_0}.
\ee
The rest is just a matter of computing the inverse of $\cal A$, whose explicit form is tridiagonal,
\be
{\cal A} = \begin{pmatrix}
-\upsilon _0 & \delta_1 & 0 & \ldots & \ldots & 0 \cr
\upsilon _0 & -\upsilon _1-\delta_1 & \delta_2 & \ldots & \ldots & 0 \cr
0 & \upsilon _1 & -\upsilon _2-\delta_2 & \delta_3 & \ldots & 0 \cr 
0 & 0 & \upsilon _2 & \ldots & \ldots & \ldots \cr
\ldots & \ldots & \ldots & \ldots & \ldots & \delta_{n-1} \cr
0 & 0 & 0 & \ldots & \ldots & -\upsilon _{n-1}-\delta_{n-1} \cr
\end{pmatrix}.
\ee

For $i_0=0$ (starting site is the origin, the left end of the chain), the entries ${\cal A}^{-1}_{i,0}$ have a relatively simple form. We find
\be
{\cal T}_{0}(i) = -({\cal A}^{-1})_{i,0} = {1 \over \upsilon _i} \; \sum_{\ell=0}^{n-1-i} \: {\delta_{i+1} \ldots \delta_{i+\ell} \over \upsilon _{i+1} \ldots \upsilon _{i+\ell}}, \qquad i=0,1,\ldots,n-1,
\label{general}
\ee
where the summand for $\ell=0$ is taken to be 1. We deduce in particular the relation
\be
\delta_i \, {\cal T}_0(i) = \upsilon _{i-1} \, {\cal T}_0(i-1) - 1.
\ee

Let us mention a few special cases. In the symmetric case $\delta_i=\upsilon _i$ for $1 \le i \le n-1$ (left and right moves have equal probabilities), we find the simple expression
\be
{\cal T}_{0}(i) = \frac{n-i}{\upsilon _i}, \qquad i=0,1,\ldots,n-1.
\ee
The ratio $ \frac{\delta}{\upsilon } $ is equal to the previous ratio
$\kappa = \frac{d}{u}$ from which the temporal aspect disappears. Hence, in the homogeneous case, namely when all $\upsilon _i$ and all $\delta_i$ are equal to $\upsilon $ and $\delta$ respectively, the formula (\ref{general}) simplifies to
\be
{\cal T}_{0}(i) = {1 \over \upsilon } \; (1+\kappa+\kappa^2+ \ldots+\kappa ^{n-1-i}) = {1 \over \upsilon } \; \frac{1-\kappa^{n-i}}{1-\kappa} , \qquad i=0,1,\ldots,n-1,
\ee

Fig.A3 shows a discrete simulation of the number of returns to state 0 which accumulate nonlinearly in time in the case of a slightly backward finite random walk.

\begin{center}
\includegraphics[width=12cm]{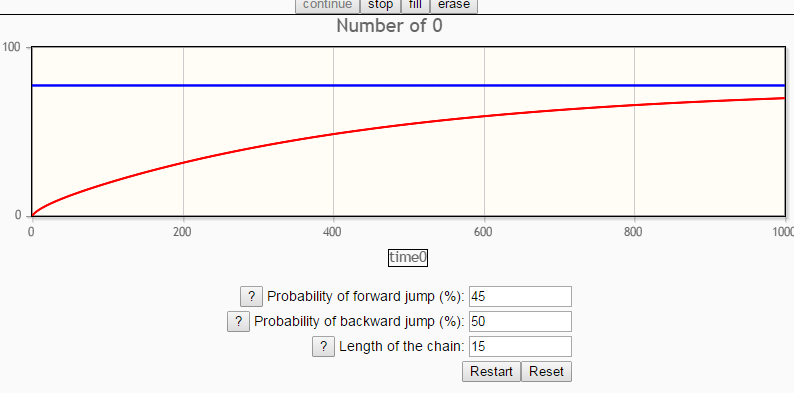} \\
\end{center}
\begin{small} \textbf{Figure A3.} Screen capture of the stochastic simulation applet showing the evolution of the number of returns to 0 for a 15-step single walk weakly backward with backward transitions 5\% more probable than forward steps. A mean number of returns to zero of 77.14 is expected for this set of values. The expected time of absorption is 471.39, with a standard deviation equal to 423.47. \end{small} \\

\end{document}